\documentclass{ws-p8-50x6-00}
\usepackage{epsfig}

\newenvironment{Pitemize}
  {\begin{list}{}%
    {\setlength{\topsep}{0mm}%
     \setlength{\parsep}{0mm}%
     \setlength{\itemsep}{0mm}%
     \setlength{\parskip}{0mm}%
    }%
  }%
  {\end{list}}

\def\bu{\bullet}
\def\ifmath#1{\relax\ifmmode #1\else $#1$\fi}%

\def\rd{\ifmath{{\mathrm{d}}}}
\def\rl{\ifmath{{\mathrm{l}}}}

\def\rp{\ifmath{{\mathrm{p}}}}
\def\ro{\ifmath{{\mathrm{o}}}}

\def\rp{\ifmath{{\mathrm{p}}}}

\def\rs{\ifmath{{\mathrm{s}}}}

\def\rT{\ifmath{{\mathrm{T}}}}

\begin{document}

\title{HBT in Relativisitic Heavy Ion Collisions}

\author{Michael Murray }
\address{Cyclotron Institute, Texas A\&M University}

\maketitle
\abstracts{
A summary of current interferometry data in relativistic heavy ions
is presented. At $\sqrt{s_{nn}}=17$GeV a sudden increase
in the pion source volume is observed for central PbPb collisions.
 This seems to imply that the pion phase density has reached a limit.
 The source size of 
different particles decreases with mass when the transverse velocity is
held constant but increases with mass when $m_\rT$ is held constant.
The antiproton source radius is larger than the proton source radius.
So far no long lived source has been seen.
The pion source size varies slowly with rapidity but more rapidly with
$m_\rT$ implying strong transverse flow. There is very slow increase 
of pion radii with $\sqrt{s}$.
}

\section{Introduction}

The goal of studying relativistic heavy ion collisions is to heat and 
compress nuclear matter to such an extent that it undergoes a phase 
transition to a quark gluon plasma. 
Such a phase transition should also make a significant difference in 
the Equation of State of nuclear matter and may therefore reveal itself 
in the space-time evolution of the source. 
Hanbury Brown and Twiss 
interferometry is sensitive to this evolution and 
can thus play a crucial role in searching for a phase transition. 
Multi-dimensional correlation function analysis allows extraction of 
the duration of particle emission. 

With HBT we hope to answer the following questions:
\begin{Pitemize}
\item[$\bu$] {Duration of Freezeout:}
Can we see a long lived source?
For a mixed phase of QGP and Hadron gas we might 
hope to see a long lived
duration of emission via a difference in the 
outward and sideward radii.\cite{QGPinHBT}
\item[$\bu$] {Azimuthal Flow:} What is the source shape in the reaction plane?
\item[$\bu$] {Multiplicity Dependence:} 
Is there a critical multiplicity above which source size 
increases rapidly?
\item[$\bu$] {Rapidity, $\sqrt{s}$, $m_\rT$ \& $\gamma_\rT$ dependence:}
Have we produced a boost invariant source?
Is there a critical $\sqrt{s}$? Can we see transverse flow?
\item[$\bu$] {3-Particle Correlations:} Is the source chaotic?
\item[$\bu$] {Phase Densities:} Have we saturated phase space?
\end{Pitemize}
This paper is an attempt to answer such questions.

\section{Two-Particle Correlations}
For a set of events that are azimuthally symmetric,
the correlation function $C_2$ is often fitted with the following 
three-dimensional Gaussian parameterization:
\begin{equation}
   C_2=1+\lambda~\exp(-R^2_{\rs}Q^2_{\rs}-R^2_{\ro}Q^2_{\ro} 
-R^2_{\rl}Q^2_{\rl}).
   \label{eq:c2}
\end{equation}
The momentum difference $\vec{Q}(=\vec{p_1}-\vec{p_2})$ of the particle 
pair is resolved into three dimensions;
$Q_\rl$ parallel to the beam axis; 
$Q_{\ro}$ is parallel to the sum
of transverse momentum of particle pairs and
and $Q_{\rs}$ which is perpendicular
to $Q_\rl$ and $Q_{\ro}$.
Typically the Longitudinal Center of Mass System (LCMS) 
is chosen as reference frame ($p_{z1}+p_{z2}=0$).

\subsection{Rapidity and $\protect m_\rT$ Dependence}
The NA49 collaboration has measured pion radii over a wide range of 
rapidity and transverse mass $m_\rT$.
Their results are shown in 
Fig.~1 and are in good agreement with those of 
NA44.\cite{NA44pbpi} 
At $\sqrt{s_{nn}}=17$GeV, the transverse radii $R_\rs$ and $R_\ro$ 
are boost invariant while
in the longitudinal direction $R_\rl$ falls as one moves away from central
rapidity. The radii decrease as $k_\rT=m_\rT-m_\pi$ increases, which is 
suggestive of a hydrodynamic expansion.\cite{CSOR94A,WEID99A}
In such a system, the HBT source radii are 
``lengths of homogeneity'' that are set by velocity and/or temperature
gradients in the fluid.

\begin{figure}
\begin{minipage}[t!]{5.5cm}
\epsfxsize=6.5cm\epsfbox{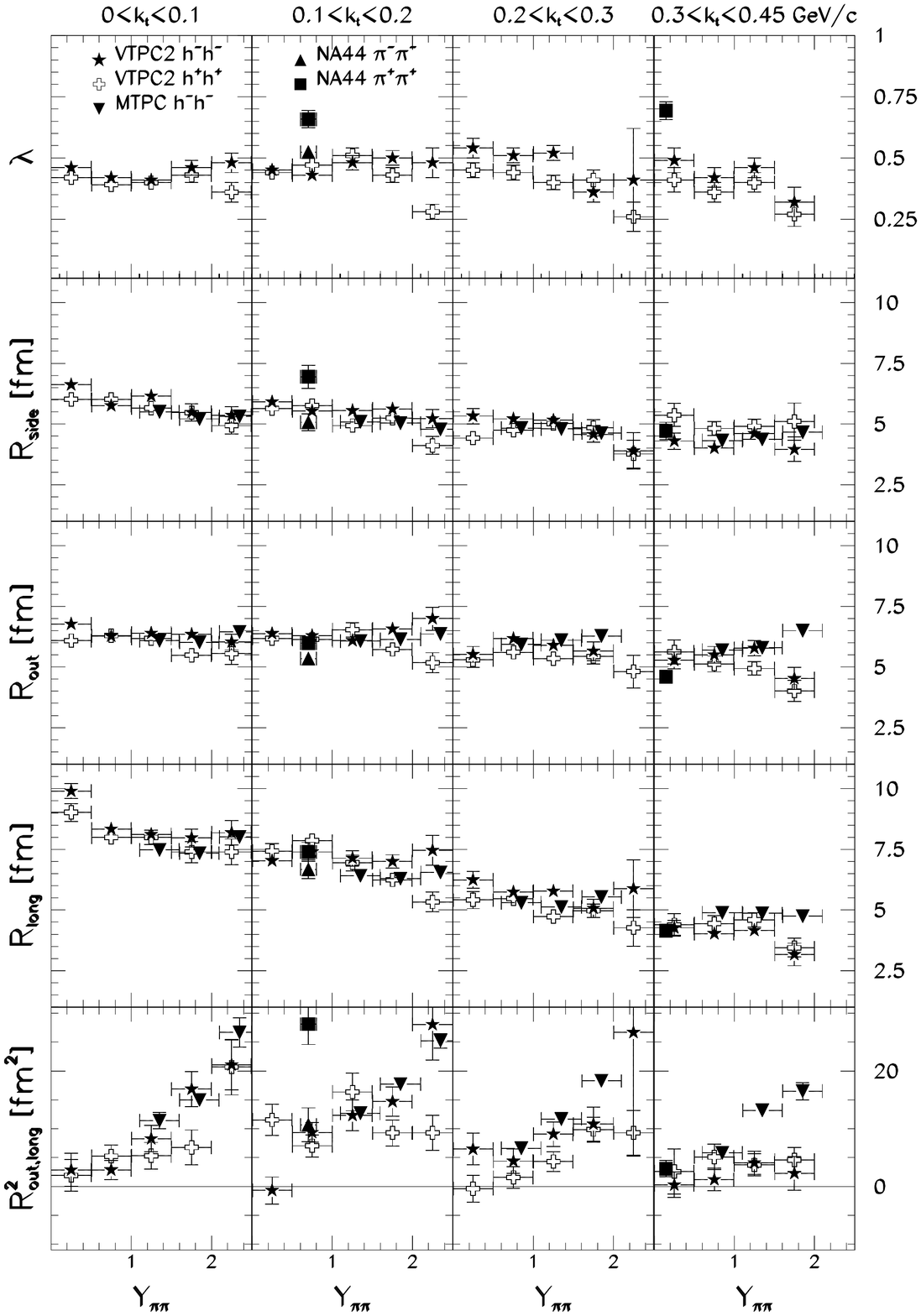}
\label{fg:NA49RvY}
\end{minipage}
\hskip1cm
\begin{minipage}[t!]{5.3cm}
 \mbox{\epsfxsize=5.8cm\epsffile{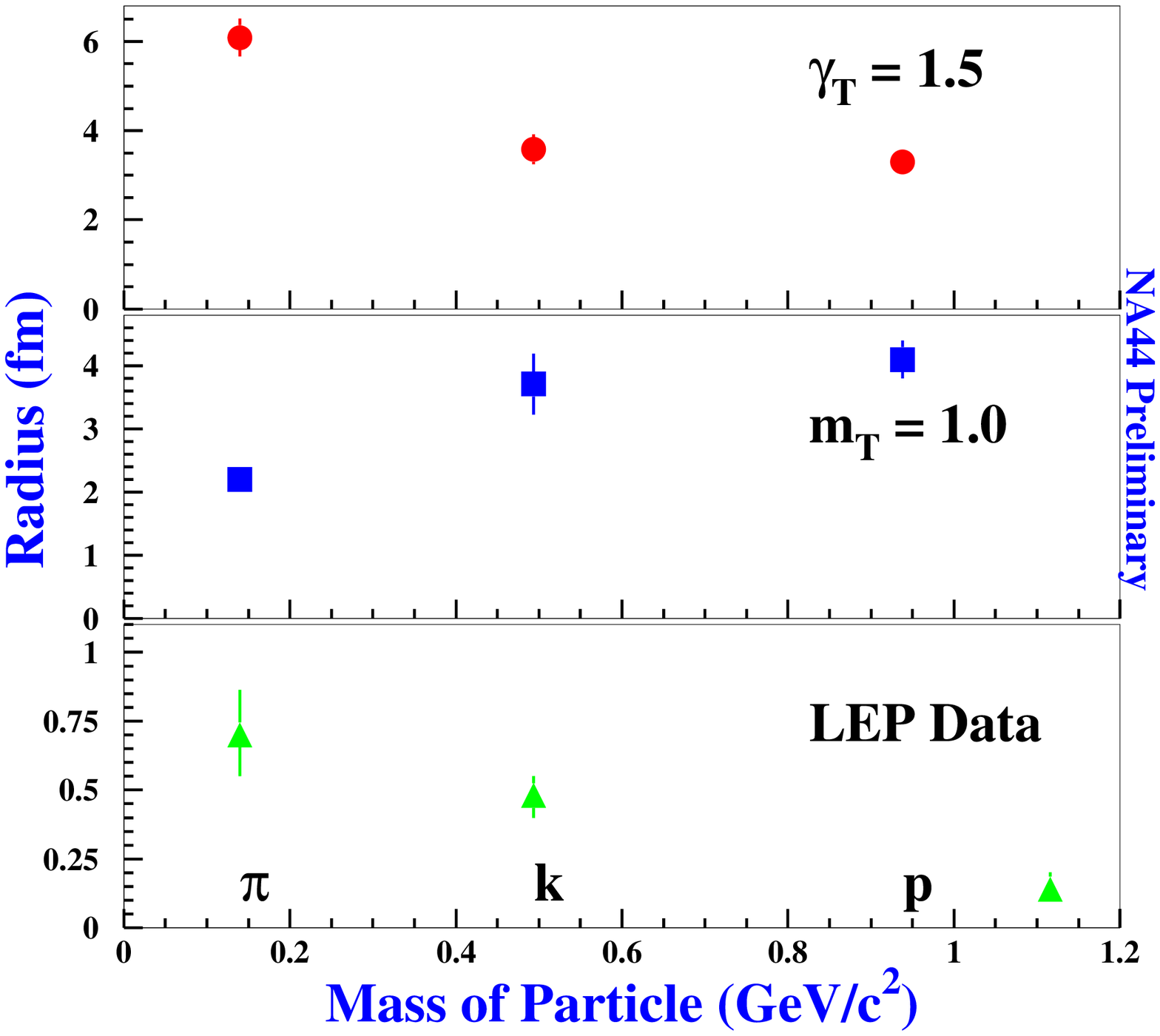}}
 \label{fg:pikp_radii}
\vskip 2mm
{\small Figure 2. Source radii versus mass for different particles.\par}
\vskip 1cm
{\small Figure 1. Radii versus rapidity and $k_\rT=m_\rT-m$ for PbPb.\par}
\vskip 5mm~
\end{minipage}
\vskip-5mm
\end{figure}

Figure~2 shows
source radii for different particles versus $m_\rT$, $\gamma_\rT$ the
Lorentz factor in the transverse direction, and for LEP data the
mass of the particle. At fixed $\gamma_\rT$ the source radius decreases
with increasing mass of the particle, while at fixed $m_\rT$ the radii increase
with mass. This could imply that the lighter pions leave the source
before the protons have frozen out.

\subsection{Energy and Multiplicity Dependence}

\begin{figure}
\begin{minipage}[t]{5.5cm}
\epsfxsize=6cm\epsfbox{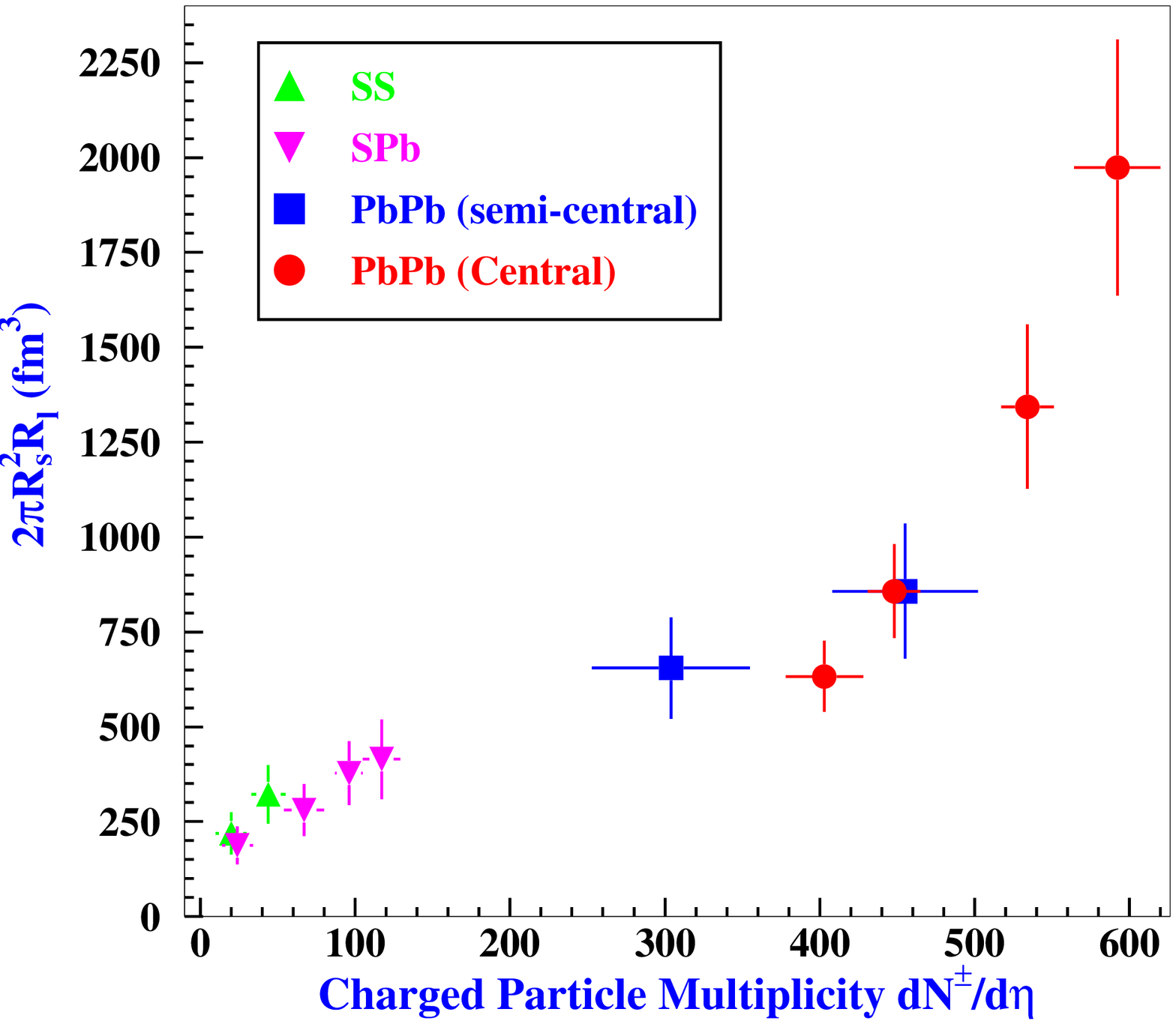}
{\small
Figure 3. Volume of the $\pi$ source versus multiplicity at
$\protect{\sqrt{s_{nn}}}\approx 17$GeV.\protect\cite{NA44VOL}
\label{fg:VolvMult}\par}
\end{minipage}
\hskip5mm
\begin{minipage}[t]{5.5cm}
\epsfxsize=6cm\epsfbox{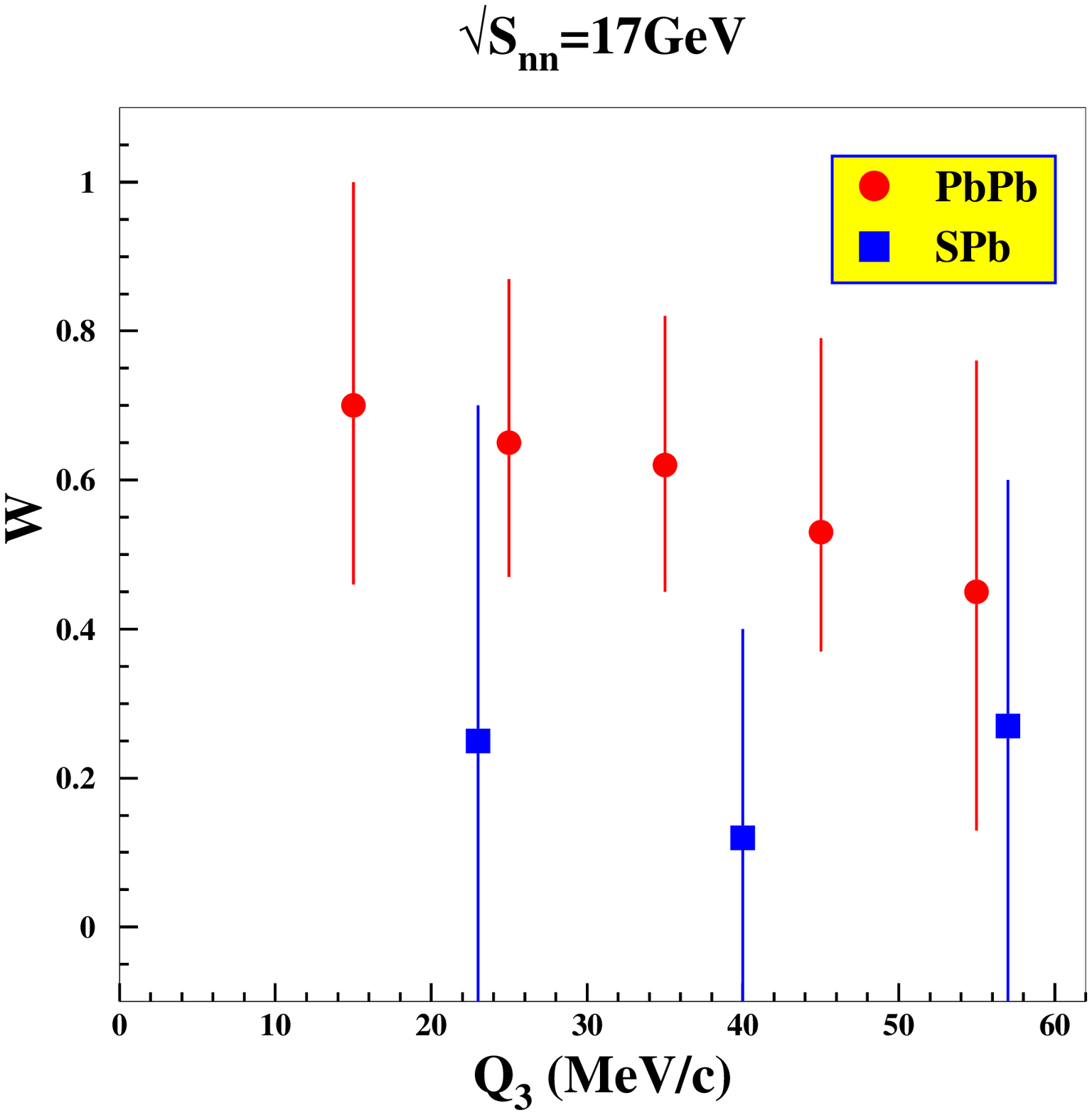}
{\small Figure 4. The strength of the $3\pi$ correlation function versus 
Q3.\protect\cite{NA443pi,WA983pi}\par}
\label{fg:pi3sps}
\end{minipage}
\vskip-5mm
\end{figure}

At the AGS $\sqrt{s_{nn}}=4.9$ GeV, the pion radii
increase as the impact parameter decreases,
and the multiplicity increases.\cite{E895}
This is also true at CERN energies.
Figure~\ref{fg:VolvMult} shows a measure of the ``homogeneous volume"
for pions versus the multiplicity of the events at
$\sqrt{\rs_{nn}}\approx 17$GeV.
 As expected, the
source volume increases with increasing multiplicity.
 For SA and peripheral
PbPb collisions, the volume rises more slowly than the multiplicity,
implying an increase in the phase space density.
However, for central PbPb
collisions there is a sharp increase in the slope of the curve.
 This can be interpreted
as the phase space density reaching some limit. It will be interesting
to see if this limit holds at the much higher multiplicities achieved
at RHIC.
At the same time, the duration of pion emission
$\delta\tau \approx \sqrt{R_\ro^2-R_\rs^2}/\beta_\rT$,
seems to
saturate with increasing multiplicity at about 4fm.

One of the most interesting results
of this conference is that STAR has found that the pion radii increase
by less than $20\%$ while $\sqrt{s}$ increases by a factor of 
7.5.\cite{STARISMD} This may be due to very rapid expansion of the source.

\subsection{Sizes in and out of the Reaction Plane}
E895 has measured the source size with respect to the reaction 
plane,\cite{E895} and has found that for semi-central collisions
$R_\rs$ is larger along the reaction plane than perpendicular
to it. 

\section{Three-Particle Correlations}
Two-particle interferometry is unable to provide a measurement of
the phase of the source functions. However, this information can
be deduced from three particle correlations,\cite{Theory3pi} if 
the emission is fully chaotic.
This phase reflects asymmetries in the source which may be
induced by geometry, flow, or resonance decays.
If the source is not completely chaotic,
the interpretation is more difficult.
One can measure the strength of the 3-particle correlation by defining
\begin{equation}
W \equiv \frac{\{C_3(Q_3)-1\}-\{C_2(Q_{12})-1\}-\{C_2(Q_{23})-1\}-\{C_2(Q_{31})-1\} }
{2\sqrt{\{C_2(Q_{12})-1\}\{C_2(Q_{23})-1\}\{C_2(Q_{31})-1\}}}
\label{eq:phase1}
\end{equation}
For a fully chaotic system we expect $W=1$. Figure~\ref{fg:pi3sps} shows
$W$ versus $Q3$ for SPb and PbPb for $\sqrt{s_{nn}} \approx 17$GeV.
For PbPb the system is more chaotic than for SPb.

\section{Phase Space Density}

A particle's phase space density is defined as
\begin{equation}
 f({\bf p}, {\bf x}) \equiv \frac{(2\pi \hbar c)^3}{(2s+1)}
  \frac{\rd^6N}{\rd p^3\rd x^3}\ \ ,
 \label{eq:fdefine}
\end{equation}
where $s$ is the particle's spin.
Averaged over the ``homogeneous" volume,
$f_\pi$ can be derived from
the ratio of the single-particle spectrum to the volume as measured by HBT, 
\cite{WEID99A,BERTSCH,E87797}
\begin{equation}
 \langle f_\pi \rangle  =
\frac{\pi^{\frac{3}{2}} (\hbar c)^3}{(2s+1)}
\sqrt{\lambda} {\rd^3N_\pi\over \rd p^3} \frac{1}{R_\rs\sqrt{R^2_\ro R^2_\rl 
- R^4_{\ro\rl}}}\ .
 \label{eq:pifaze}
\end{equation}

\begin{minipage}[t]{5.3cm}
\epsfig{file=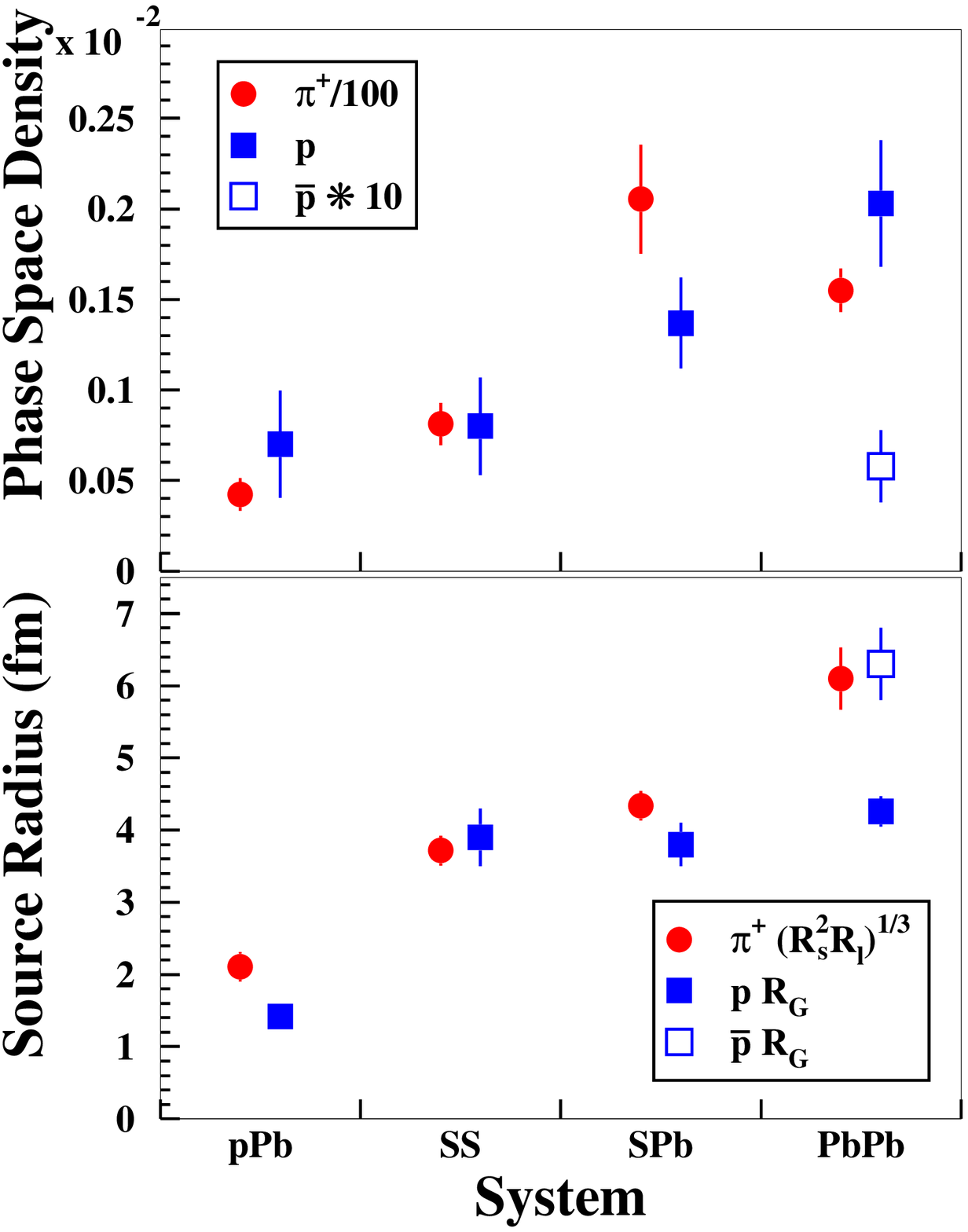,width=5.cm}
\vskip 1mm
{\small Figure 5. Phase space densities $ \langle f \rangle $ and 
source radii  for $\pi^+$ and p and  $\bar \rp$.\par}
\end{minipage}
\hskip6mm
\begin{minipage}[t]{5.cm}
\vskip-6cm
Figure~5 shows the system dependence of the
phase space densities and source radii for $\pi^+$, p and $\bar \rp$.
The $\pi^+$ and p phase space densities generally
increase with system size.
At $\sqrt{s_{nn}}=17.3$GeV
$$\langle\! f_{\bar \rp}\!\rangle\! \ll\! \langle\! f_\rp\! \rangle\! \ll
\langle\! f_{\pi^+}\! \rangle\! <\! \langle\! f_{\pi^-}\! \rangle\! \ll 1.$$ 
The large antiproton radius may be, because only $\bar \rp$s emitted
from the surface of the source can avoid annihilation.\cite{MROWCZYN}
This would imply that the observed  $\bar \rp$'s would have
a larger RMS freeze-out radius than the protons.
\end{minipage}

\section{Conclusions}
Interferometry is now a mature field able to make detailed 
observations of the hadronic source. It is now possible to measure the
extra width of the source in the reaction plane and the duration of
the pion emission. At $\sqrt{s_{nn}}=17$GeV, we see a sudden increase
in the pion source volume above a certain multiplicity. This seems to 
imply that the pion phase space has reached some kind of limit. It will
be interesting to see if this limit is broken at RHIC. At 
$\sqrt{s_{nn}}=17$GeV 
collisions, the duration of pion emission rises with multiplicity and then
saturates at 
$\delta\tau \approx 4$fm. So far, no long lived source has been seen.
The pion source size varies slowly with rapidity but more rapidly with
$m_\rT$, implying strong transverse flow. 
It is not clear
how to compare radii from different particles. Should we study them
at the same transverse velocity  or the same transverse mass?
The antiproton radius looks larger than the proton radius. However, this 
may be due to annihilation in the interior of the source.
The first data from STAR show 
radii very similar to those seen at lower energies.
  In looking to the future we should draw inspiration from the past.
Hanbury Brown and Twiss invented the interferometry technique to measure
the size of stars using the interference of photons. Perhaps the next
great advance in our field will come from photon interferometry of 
heavy ion collisions. This would allow us to ``see" the beginning of 
these collisions and not just their end.

\section*{Acknowledgments}
I would like to thank the organizers of ISMD for their invitation and 
generous
support. I also would like to thank the  many representatives of experiments for giving me plots.
This work was supported by the DOE under contract FG03-93-ER40773.

\end{document}